\documentclass[preprint]{elsarticle}

\usepackage{lineno}

\usepackage[%
a4paper=true,%
breaklinks=true,%
colorlinks=true%
]{hyperref}

\usepackage{color}
\usepackage{bm}
\usepackage{amsmath}
\usepackage{amssymb}
\usepackage{graphicx}
\usepackage{subfigure}
\usepackage{subfigmat}
\usepackage{multirow}
\usepackage{algorithm}
\usepackage{algorithmic}

\journal{Acta Astronautica}

\begin{document}

\begin{frontmatter}

\title{Hybrid SGP4 orbit propagator}

\author[GRUCACI_CCT]{Juan F\'elix San-Juan\corref{mycorrespondingauthor}}
\address[GRUCACI_CCT]{Scientific Computing Group (GRUCACI), University of La Rioja, Madre de Dios 53, 26006 Logro\~no, Spain}
\cortext[mycorrespondingauthor]{Corresponding author. Tel.: +34 941299440;  fax: +34 941299460.}
\ead{juanfelix.sanjuan@unirioja.es}

\author[GRUCACI_DEP]{Iv\'an P\'erez}
\address[GRUCACI_DEP]{Scientific Computing Group (GRUCACI), University of La Rioja, San Jos\'e de Calasanz 31, 26004 Logro\~no, Spain}
\ead{ivan.perez@unirioja.es}

\author[GRUCACI_UGR]{Montserrat San-Mart\'in}
\address[GRUCACI_UGR]{Scientific Computing Group (GRUCACI), University of Granada, Santander 1, 52005 Melilla, Spain}
\ead{momartin@ugr.es}

\author[GRUCACI_DEP]{Eliseo P. Vergara}
\ead{eliseo.vergara@unirioja.es}

\begin{abstract}
Two-Line Elements (TLEs) continue to be the sole public source of orbiter observations. The accuracy of TLE propagations through the Simplified General Perturbations-4 (SGP4) software decreases dramatically as the propagation horizon increases, and thus the period of validity of TLEs is very limited. As a result, TLEs are gradually becoming insufficient for the growing demands of Space Situational Awareness (SSA). We propose a technique, based on the hybrid propagation methodology, aimed at extending TLE validity with minimal changes to the current TLE-SGP4 system in a non-intrusive way. It requires that the institution in possession of the osculating elements distributes hybrid TLEs, HTLEs, which encapsulate the standard TLE and the model of its propagation error. The validity extension can be accomplished when the end user processes HTLEs through the hybrid SGP4 propagator, HSGP4, which comprises the standard SGP4 and an error corrector.
\end{abstract}

\begin{keyword}
Artificial satellite theory \sep Orbit propagator \sep Hybrid propagation methodology \sep SGP4 \sep TLE \sep SSA
\end{keyword}

\end{frontmatter}

%\linenumbers

\section{Introduction}
Space Situational Awareness (SSA) is one of the main issues that affect the space sector nowadays. The huge amount of debris in orbit around the Earth, especially concentrated in the most populated regions, LEO and GEO, pose an unprecedented hazard on space assets and, consequently, on the services they provide to the society. The magnitude of the problem has led the authorities to introduce regulations on the space activity, in such a way that no launching permission is granted unless certain conditions are fulfilled. Those conditions include proving that the space objects will not be invading some protected regions during a fixed time span, which can extend even to a hundred years. Suddenly, the time horizon of the propagations to be computed has increased significantly.

In addition, the need for the accurate propagation of thousands of catalogued objects has arisen in order to allow for the early detection of probable collisions with operative satellites. Consequently, a new need has been imposed on orbit propagators: the computing speed. As a result, we have recently seen a renewed interest in the orbit propagation methods, with a clear demand for enhancements, both in terms of accuracy and computing efficiency.

The problem of orbit propagation consists in determining the position and velocity of an orbiter at a given future time $t_f$ from its initial conditions at an instant $t_i$. In general, it is a non-integrable problem which implies handling a three second-order or six first-order differential-equation system.

Three well-known techniques have been classically considered in order to solve this problem. They can be characterized according to the perturbation model taken into account, the formulation of the equation of motion, and the integration method applied to its resolution, which can be analytical or numerical.

\textit{General perturbation theories} generate an analytical solution through the application of perturbation methods \cite{bro1959_astnodrag, koz1962_2ordnodrag, lyd1963_smalleccinclbrow, aks1970_order2, kin1977_ord3ast}. The result is an explicit function of time and some physical constants, which implies very fast evaluations, although the accuracy can be reduced due to the fact that, in order to avoid cumbersome expressions, the analytical solution is usually a low-order approximation in which only the most relevant forces are considered.

\textit{Special perturbation theories}, on the contrary, are characterized by integrating the differential equations numerically \cite{kin1989_numint, lon1989_gtdsmaththeor1}. This technique is conducive to higher accuracy because it allows taking into account complex perturbation models, although the need for small integration steps translates into long computational time.

\textit{Semianalytical techniques} constitute a combined approach \cite{liu1980_sstclosearth, nee1998_currentdsst, cef2010_accessdsst}. Complex perturbing effects can be included in the formulation, which is then transformed through analytical methods in order to remove the short-period components. The slower dynamics of the remaining components allows performing a numerical integration through longer integration steps. As a result, a compromise between accuracy and agility can be achieved.

Different attempts to enhance these classical propagation methods have focused on approximating the physical models more accurately to the real phenomena, particularly the atmospheric drag \cite{ker2015_genpert4lt} and the solar radiation pressure \cite{mon2015_srpgalileo, guo2016_t20srpcompass}, since non-gravitational perturbations are especially challenging to model.

In the case of numerical integration, the emphasis has been placed on achieving both a higher accuracy and faster algorithms, which can take advantage of parallel computing \cite{sha1980_parallelorbdet_phd} in multicore and GPU computational systems. Alternatively, the research activity on analytical and semianalytical techniques has mainly been focused on extending to higher orders the required series expansions \cite{san2011gru_ppkbz9}, as well as on completing their perturbation models.

More recently, the \textit{hybrid propagation methodology}, aimed at improving the accuracy of any orbit propagator, has been proposed. This methodology can combine any of the three classical integration methods with a forecasting technique. The aforementioned simplifications and inaccuracies in the integration methods lead to solutions that are approximate, and consequently can be improved. In order to achieve it, forecasting techniques based on either statistical time series models \cite{san2012gru_sarimahop, san2014gru_montse_phd, san2016gru_tshop} or machine learning methods \cite{per2013gru_nnhop, per2015gru_ivan_phd} can be used. They can model the dynamics of the difference between the integrated approximate solution and the real behavior, during an initial control interval, with the aim of reproducing it later, when real ephemerides are no longer available. That error forecast represents a correction that, once added to the approximate solution, generates the improved final result. In conclusion, the hybrid propagation methodology constitutes a non-intrusive technique that can be applied to the enhancement of any orbit propagator.

The time series of the error during the initial control interval always show a systematic pattern that repeats in every orbiter revolution; depending on each case, a trend, usually linear, can also exist. Consequently, methodologies based on decomposition are best suited to modeling these error time series. Exponential smoothing techniques, as well as Box-Jenkins methodology, are the most important decomposition methods. The latter was used in a previous study \cite{san2012gru_sarimahop}, although we subsequently adopted the former because of its greater simplicity, speed and predictive capability. The classical Holt-Winters method \cite{win1960_forecexpma}, in its additive version, is the exponential smoothing technique that has yielded the best results thus far. Nevertheless, an ongoing line of research is the formulation of Holt-Winters methodology in the state space, which allows modifying and increasing its components, assigning a probability distribution, as well as obtaining maximum likelihood estimates.

Concerning machine learning methods, some architectures of artificial neural networks have been explored, although this line of research is not so advanced as the statistical time series one. The application of other neural network configurations to the hybrid methodology for orbit propagation, as well as alternative machine learning techniques, is currently being analyzed.

Nowadays, the sole universal source of observations available for the use of the SSA community is distributed as Two-Line Elements (TLEs),\footnote{\url{http://www.space-track.org/}} which are generated as mean elements based on the Kozai mean motion \cite{koz1962_2ordnodrag}. The propagation of TLEs needs to be done through the Simplified General Perturbations-4 (SGP4) software \cite{hoo1980_spacetrack3, val2006_spacetrack3rev}, which is the propagator specially adapted to TLE specifications, although it only considers the main perturbing effects. Nevertheless, no other propagator is recommended to be used with TLEs. In addition, since TLEs are based on mean elements instead of osculating elements, the associated uncertainty causes a rapid loss of accuracy, and therefore a very reduced propagation horizon. As a result, the current system based on TLEs and SGP4 is becoming insufficient for SSA growing demands.

We propose a non-intrusive enhancement of the TLE-SGP4 system based on the hybrid propagation methodology: the HTLE-HSGP4 system. It requires the modeling of the error associated to the SGP4 propagation of a TLE during an initial control interval, which should be done by the TLE distributor, followed by the inclusion of some parameters that define the deduced error model in the distributed hybrid TLE, that is, the HTLE. The end user can benefit from an extension in the HTLE validity if the propagation is done through the hybrid SGP4 propagator, that is, HSGP4, which comprises both the standard SGP4 and an error corrector that calculates the correction to apply to the SGP4 output from the additional parameters included in the HTLE.

The outline of this paper starts with a succinct formal description of the hybrid propagation methodology in Section 2. Next, Section 3 provides a general overview of the Holt-Winters algorithm implemented for the modeling of the SGP4 error and the generation of the corresponding correction. Then, the proposed hybrid HSGP4 orbit propagator is presented in Section 4 by means of a real example, the sun-synchronous satellite \textit{Deimos~1}. Finally, the main concepts of the study are summarized in Section 5.

\section{Hybrid propagation methodology}

Similar to any other propagation theory, the hybrid propagation methodology is designed to determine an estimation $\hat{\mathbf{x}}_f$ of the position and velocity of an orbiter at a final instant $t_f$, from its position and velocity $\mathbf{x}_1$ at an initial instant $t_1$, where $\mathbf{x}$ represents the complete set of six variables, which can be referred to any canonical or non-canonical coordinate system.

In the first place, an initial approximation to $\hat{\mathbf{x}}_f$ must be calculated through the application of the integration method $\mathcal{I}$, which represents any of the aforementioned three classical theories, to the initial conditions $\mathbf{x}_1$:

\begin{equation}\label{sol_int}
\mathbf{x}_f^{\mathcal{I}} = \mathcal{I}(t_f,\mathbf{x}_1).
\end{equation}\par

In general, the integration method $\mathcal{I}$ includes some simplifications so as to make the process viable and affordable, and thus $\mathbf{x}_f^{\mathcal{I}}$ is just an approximation of $\hat{\mathbf{x}}_f$. A second step, in which an estimation of the difference is predicted, is therefore necessary. In order to achieve it, a forecasting technique has to model the dynamics missing from the approximation generated by $\mathcal{I}$. For that purpose, a set of either real observations or accurately generated ephemerides $\mathbf{x}$, which represent the real dynamics of the orbiter, must be available during an initial \textit{control interval} $[t_1,t_T]$. By means of those values, the error of the integration method, that is, its difference with respect to the orbiter real behavior, can be determined for any instant $t_i$ in the control interval as

\begin{equation}\label{err_ti}
\mathcal{\bm\varepsilon}_i = \mathbf{x}_i - \mathbf{x}_i ^{\mathcal{I}},
\end{equation}

\noindent
where $\mathbf{x}_i$ represents the accurate ephemeris and $\mathbf{x}_i ^{\mathcal{I}}$ the approximation generated by the integration method $\mathcal{I}$.

The six time series of $\mathcal{\bm\varepsilon}_i$ values from $t_1$ to $t_T$, $\mathcal{\bm\varepsilon}_1,\ldots,$ $\mathcal{\bm\varepsilon}_T$, which we call \textit{control data}, contain the dynamics to be modeled and reproduced by the forecasting technique. After the adjustment process, an estimation of the error at the final instant $t_f$, $\hat{\mathcal{\bm\varepsilon}}_f$, can be determined, thereby allowing for the calculation of the desired value of $\hat{\mathbf{x}}_f$ as

\begin{equation}\label{sol_hyb}
\hat{\mathbf{x}}_f = \mathbf{x}_f^{\mathcal{I}} + \hat{\mathcal{\bm\varepsilon}}_f.
\end{equation}

\section{Exponential smoothing method: Holt-Winters algorithm}

The modeling of the time series of the error during the initial control interval can be done with several techniques; we have chosen the Holt-Winters algorithm not only because of its simplicity and good results, but also due to the portability of the developed model, which constitutes a desirable quality in order to allow the model to be integrated into the extended TLEs in a compact way.

Holt-Winters \cite{win1960_forecexpma} is one of the so-called exponential smoothing methods. They consider that a time series $\varepsilon_t$ is composed of trend $\mu_t$, or secular variation, periodic oscillation $S_t$, or seasonal component, and unpredictable or random variation $\nu_t$. In the case of an additive composition, those three components add up. In particular, the Holt-Winters algorithm considers a linear trend $\mu_t=A+B t$ with level $A$ and slope $B$. Therefore, the value of the time series at an instant $t$ can be estimated as

\begin{equation} \label{SE1}
\hat{\varepsilon}_{t}=A_{t-1}+ B_{t-1}+S_{t-s},
\end{equation}

\noindent
where $s$ represents the period of the seasonal component.

Then, $A$, $B$, and $S$ can be determined for the same instant $t$ as the weighted sum of two values, one based on the time series real value, $\varepsilon_t$, which must be known during the control interval, and the other dependent on the previous estimation of the time series components,

\begin{eqnarray} 
	A_t & = & \alpha (\varepsilon_t-S_{t-s}) + (1-\alpha) (A_{t-1}+B_{t-1}), \nonumber\\[.5ex]
	B_t & = & \beta(A_t-A_{t-1}) + (1-\beta) B_{t-1},\label{SE2}\\[.5ex]
	S_t  & = & \gamma(\varepsilon_t-A_{t}) + (1-\gamma) S_{t-s}.\nonumber
\end{eqnarray}

\noindent
The three weights, $\alpha$, $\beta$, and $\gamma$, with values in the interval $[0,1]$, are known as \textit{smoothing parameters}.

\begin{algorithm}
\begin{algorithmic}[1]
\REQUIRE $s$, $c$, $h$, and $\{\varepsilon_t\}_{t=1}^{T}$
\ENSURE $\hat{\varepsilon}_{T+h|T}$
\STATE Estimate the values of $A_0, B_0,S_{-s+1},\ldots, S_{-1},S_0$
\FOR {$t=1;\,t\leq T;\,t=t+1$}
\STATE $A_t = \alpha (\varepsilon_t-S_{t-s}) + (1-\alpha)(A_{t-1}+ B_{t-1})$
\STATE $B_t = \beta(A_t-A_{t-1}) + (1-\beta) B_{t-1}$ 
\STATE $S_t = \gamma(\varepsilon_t-A_{t}) + (1-\gamma) S_{t-s}$
\STATE $\hat{\varepsilon}_{t} = A_{t-1}+ B_{t-1} + S_{t-s}$
\ENDFOR
\STATE Select \texttt{error$\_$measure} $\in$ \{MSE, MAE, MAPE\} and express it as a function of the smoothing parameters
\STATE Obtain the smoothing parameters that minimize \texttt{error$\_$measure} using the L-BFGS-B method
\STATE Calculate $A_T, B_T, S_{T-s+1},\ldots, S_{T-1}, S_T$ for the optimum smoothing parameters
\STATE  $\hat{\varepsilon}_{T+h|T} = A_T + h B_T + S_{T-s+1+h\,\mathrm{mod}\, s}$
\RETURN  $\hat{\varepsilon}_{T+h|T}$
\end{algorithmic}
\caption{Holt-Winters}\label{algHW}
\end{algorithm}

The algorithm~\ref{algHW} \cite{san2014gru_montse_phd}, which we have implemented in the R statistical programming language \cite{r2015_sw_r}, shows how to apply the Holt-Winters method to the prediction of future time series values. The inputs to the algorithm are the amount of data per revolution, $s$, the number of revolutions in the control interval, $c$, the number of time steps after the control interval for which the time series value has to be predicted, $h$, and the control data, $\left\{\varepsilon_t\right\}_{t=1}^{T}$, with $T=s \times c$. The output is $\hat{\varepsilon}_{T+h|T}$, that is, the forecast of the time series at the final instant $t_f=t_{T+h}$, based on the last control data, $\varepsilon_T$.

The algorithm starts by estimating the initial parameters $A_0$, $B_0$, $S_{-s+1},\ldots,$ $S_{-1}$, and $S_0$, which is accomplished through a classical additive decomposition into trend and seasonal variation over the first orbiter revolutions. The first step in this process consists in applying a moving-average filter to the time series, which yields the trend. Then, a linear regression over the trend allows obtaining the initial values of the level $A_0$ and slope $B_0$. After that, the difference between the time series and the trend generates the seasonal component, which consists of $s$ points per revolution. Finally,  $S_{-s+1},\ldots,$ $S_{-1}$, and $S_0$ can be obtained as the $s$ average values of the seasonal-component equivalent points in each revolution.

Then, an iterative process takes place by applying Equations~\eqref{SE1} and~\eqref{SE2} to the control interval, starting from $t_1$ until $t_T$ (lines 2--7). As a result, a set of time series estimated values, $\hat{\varepsilon}_{t}$, dependent on the smoothing parameters, are determined. The minimization of the estimation error with respect to the time series real values, $\varepsilon_t$, leads to the optimal values of $\alpha$, $\beta$, and $\gamma$.

A very common optimization method among specialists in time series and machine learning is the limited memory L-BFGS algorithm, which is a quasi-Newton technique that approximates the BFGS algorithm \cite{sha1970_quasinewt}, named after Broyden, Fletcher, Goldfarb, and Shanno, using a limited amount of computer memory. The reason for its popularity lies in its reduced memory requirements, which makes it especially appropriate for complex optimization problems. The L-BFGS algorithm is usually employed to estimate parameters, although it does not allow imposing constraints on them. Another optimization algorithm called L-BFGS-B \cite{byr1995_limmemopt} extends L-BFGS to handle parameter constraints. In our particular case, the three smoothing parameters whose values have to be optimized, $\alpha$, $\beta$, and $\gamma$, must be constrained to the interval $[0,1]$, hence we choose L-BFGS-B as the optimization algorithm.

Three different error functions can be used as an optimization criterion, mean square error, MSE, mean absolute error, MAE, and mean absolute percentage error, MAPE:

\begin{eqnarray} 
\displaystyle \mbox{MSE} & = & \frac{1}{T} \sum_{t=1}^{T}{(\varepsilon_t-\hat{\varepsilon}_{t})^2}, \nonumber \\[1ex]
\displaystyle \mbox{MAE} & = & \frac{1}{T} \sum_{t=1}^{T}{|\varepsilon_t-\hat{\varepsilon}_{t}|}, \label{med.err} \\[1ex]
\displaystyle \mbox{MAPE} & = & \frac{1}{T} \sum_{t=1}^{T}{\left| \frac{\varepsilon_t-\hat{\varepsilon}_{t}}{\varepsilon_t} \right|} 100. \nonumber
\end{eqnarray}

Once the optimal smoothing parameters $\alpha$, $\beta$, and $\gamma$ have been found, the time series parameters $A_T$, $B_T$, $S_{T-s+1},\ldots,$ $S_{T-1}$, $S_T$ are determined for the last period of the control data, from which the forecast time series value at the final instant, that is, $h$ epochs ahead, $\hat{\varepsilon}_f=\hat{\varepsilon}_{T+h|T}$, can be calculated as the addition of the last control data level $A_T$, slope component $hB_T$, and equivalent seasonal value in the previous revolution $S_{T-s+1+h\,\mathrm{mod}\, s}$ (line 11).

\section{Hybrid SGP4 orbit propagator}

In this section, the hybrid propagation methodology is applied to the SGP4 orbit propagator so as to improve the perturbation model considered in its original design. The sun-synchronous orbit of the \textit{Deimos~1} Earth imaging satellite will be used to illustrate this technique. The TLE employed for this purpose is

{\small
\begin{verbatim}
1 35681U 09041A   11124.21233382  .00000325  00000-0  63164-4 0  9994
2 35681 098.0717 023.8270 0000845 081.0832 279.0474 14.69441166 94523
\end{verbatim}}

This TLE is transformed into osculating orbital elements by using SGP4, with the aim of serving as the initial conditions to be accurately propagated in order to generate a set of precise ephemerides, which we will call pseudo-observations. In this case, pseudo-observations were generated from real data by the satellite operator, Elecnor Deimos. We will use them in order to illustrate the methodology, although the process of generating them would not be necessary if real observations were available, which is the case for the TLE distributor. The perturbations taken into account in this numerical propagation include a $60\times60$ Earth gravitational potential model, atmospheric drag, luni-solar effect, solar radiation pressure including eclipses, Earth albedo, Earth infra-red (IR) re-radiation, Earth solid tides, and relativistic effect. The pseudo-observations resulting from a $30$-day numerical propagation of this model are used for two purposes: firstly, as the real ephemerides that will allow for the modeling of the SGP4 error during the initial control interval, and the subsequent generation of the parameters needed by the Holt-Winters time series forecaster; secondly, as a reference for comparisons between both versions of the SGP4 propagator: the standard and the hybrid, which we call HSGP4. This study will be conducted in terms of the Delaunay action-angle variables $(l,g,h,L,G,H)$, which are related to the classical orbital elements $(a,e,i,\Omega,\omega,M)$ according to the following expressions: $l=M$, $g=\omega$, $h=\Omega$, $L=\sqrt{\mu a}$, $G=\sqrt{\mu a (1-e^2)}$, $H=\sqrt{\mu a (1-e^2)} \cos i$.

Next, with the aim of modeling the error of SGP4 with respect to the pseudo-observations, the TLE is also propagated with SGP4 during $30$ days. Ephemerides are taken with a sampling period of ten minutes, which corresponds to approximately one tenth of the revolution time. An initial interval of $100$ samples, that is, roughly ten satellite revolutions or $0.7$ days, is selected as the control interval for the application of the algorithm described in the previous section. A detailed study about the effects of the sampling period and the control interval duration can be seen in \cite{san2014gru_montse_phd}. Finally,  a time series representing the error of SGP4 with respect to the pseudo-observations is generated, by subtracting both sets of ephemerides, for each of the six Delaunay variables, $(\varepsilon_t^l,\varepsilon_t^g,\varepsilon_t^h,\varepsilon_t^L,\varepsilon_t^G,\varepsilon_t^H)$. These are the time series which will be individually forecast using univariate Holt-Winters models.

As a first step previous to the application of the Holt-Winters algorithm, the most relevant seasonal component for each time series is determined through the analysis of periodograms and autocorrelation functions, ACF. An extensive study concerning the identification of the main periodicities can be found in \cite{san2014gru_montse_phd}. In all the cases, the Keplerian period, very near to 100 minutes, that is, ten samples, appears as the main component in the control interval being analyzed. The first six satellite revolutions are used to deduce the initial values of the level and slope of the trend $(A_0, B_0)$, as well as the ten initial points $(S_{-s+1},\ldots, S_{-1}, S_0)$, one per sample, which characterize the seasonal component, according to the additive decomposition method described in the previous section.

Then, the Holt-Winters algorithm is applied. The use of the mean square error, MSE, as the error function to be optimized usually yields the best results, and thereby is the one we select. We consider that convergence has been achieved when the MSE reaches a value under $10^{-15}$. A more detailed analysis concerning the choice of the error function can be found in \cite{san2014gru_montse_phd}.

As a result, both the trend and seasonal component that adjust optimally to the time series being modeled during the control interval can be determined. Those two components, which can be specified through twelve parameters (the level and slope of the trend, plus the ten points of the seasonal component) are obtained for the last instant in the control interval, although can be easily referred to the initial instant, for the sake of compactness, by simply changing the level. Once the six time series have been modeled $(\hat{\varepsilon}_t^l,\hat{\varepsilon}_t^g,\hat{\varepsilon}_t^h,\hat{\varepsilon}_t^L,\hat{\varepsilon}_t^G,\hat{\varepsilon}_t^H)$, the error of SGP4 can be forecast for any future instant, so that its addition to the corresponding SGP4 ephemerides constitutes the HSGP4 output. Nevertheless, it is worth noting that any effect not present during the control interval will not be reproduced in the future error forecast.

In order to validate this methodology, the initial conditions are propagated through both the standard SGP4 and HSGP4, and then their corresponding errors with respect to the pseudo-observations are  compared in terms of both position and orbital elements.

Table~\ref{tdist_rms_lghLGH} shows the RMS position errors for different propagation spans, as well as the relative improvement achieved by HSGP4 over SGP4 as a consequence of having a better force model, derived from the processing of the information contained in the pseudo-observations. As can be seen, the position error of SGP4 can be reduced by one order of magnitude after a 30-day propagation, reaching a value of only 27 km, which is the error of SGP4 for a propagation of just $3.25$ days. Similar conclusions can be drawn from Table~\ref{tvel_rms_lghLGH} for the RMS velocity errors, which remain one order of magnitude lower for HSGP4 than for SGP4, with a value of 28 m/s after a 30-day propagation.

\begin{table}[htbp!!]
%\fontsize{10}{10}\selectfont
\caption{Comparison of RMS position errors (km) of both SGP4 and HSGP4 for several propagation spans.}
\label{tdist_rms_lghLGH}
\centering 
\begin{tabular}{cccc}
\hline
Propagation span & SGP4 & HSGP4 & Improvement (\%) \\
\hline
$\phantom{}0.7$ day & $\phantom{44}5.698$ & $\phantom{4}0.441$ & $92.26$ \\[.5ex]
$\phantom{l.}1$ day & $\phantom{44}7.998$ & $\phantom{4}0.981$ & $87.74$ \\[.5ex]
$\phantom{3.}2$ days & $\phantom{4}16.369$ & $\phantom{4}2.399$ & $85.34$ \\[.5ex]
$\phantom{.3}7$ days & $\phantom{4}58.366$ & $\phantom{4}9.505$ & $83.71$ \\[.5ex]
$\phantom{.}30$ days & $270.745$ & $27.005$ & $90.03$ \\[.5ex]
\hline
\end{tabular}
\end{table}

\begin{table}[htbp!!]
%\fontsize{10}{10}\selectfont
\caption{Comparison of RMS velocity errors (m/s) of both SGP4 and HSGP4 for several propagation spans.}
\label{tvel_rms_lghLGH}
\centering 
\begin{tabular}{cccc}
\hline
Propagation span & SGP4 & HSGP4 & Improvement (\%) \\
\hline
$\phantom{}0.7$ day & $\phantom{44}6.071$ & $\phantom{4}0.430$ & $92.92$ \\[.5ex]
$\phantom{l.}1$ day & $\phantom{44}8.535$ & $\phantom{4}0.961$ & $88.74$ \\[.5ex]
$\phantom{3.}2$ days & $\phantom{4}17.479$ & $\phantom{4}2.370$ & $86.44$ \\[.5ex]
$\phantom{.3}7$ days & $\phantom{4}62.331$ & $\phantom{4}9.883$ & $84.14$ \\[.5ex]
$\phantom{.}30$ days & $289.138$ & $28.076$ & $90.29$ \\[.5ex]
\hline
\end{tabular}
\end{table}

The errors of both propagators, SGP4 and HSGP4, in terms of orbital elements, are summarized in Table~\ref{torb_lghLGH} for the same propagation spans specified for the position and velocity errors. A brief analysis of that information allows concluding that the errors are very similar for the semimajor axis and the eccentricity, whereas the behavior of the hybrid propagator is worse than the standard SGP4 for the inclination and the right ascension of the ascending node. Nevertheless, the most remarkable fact concerns the argument of the perigee and the mean anomaly, whose opposite behavior is very noticeable.

\begin{table}[htbp!!]
%\fontsize{10}{10}\selectfont
\caption{Greatest orbital element errors of both SGP4 and HSGP4 after several propagation spans.}
\label{torb_lghLGH}
\centering
\begin{tabular}{llrrrrr}
\hline
Orbital element & Propagator & $0.7$ days & $1$ day & $2$ days & $7$ days & $30$ days \\
\hline
\multirow{2}{*}{$a$ (km)} & SGP4 & $0.22$ & $0.22$ & $0.25$ & $0.42$ & $1.43$ \\[.5ex]
& HSGP4 & $-0.10$ & $-0.10$ & $0.17$ & $-0.39$ & $-1.37$ \\[1.5ex]
\multirow{2}{*}{$e$ ($\times 10^{-5}$)} & SGP4 & $-5.67$ & $-5.67$ & $-5.67$ & $6.62$ & $37.30$ \\[.5ex]
& HSGP4 & $-3.45$ & $7.34$ & $-8.21$ & $-12.98$ & $36.37$ \\[1.5ex]
\multirow{2}{*}{$i$ (deg $\times 10^{-3}$)} & SGP4 & $3.87$ & $3.87$ & $3.87$ & $3.87$ & $-4.38$ \\[.5ex]
& HSGP4 & $0.71$ & $-3.66$ & $-3.66$ & $5.21$ & $24.66$ \\[1.5ex]
\multirow{2}{*}{$\Omega$ (deg $\times 10^{-3}$)} & SGP4 & $2.93$ & $2.93$ & $3.05$ & $3.12$ & $4.02$ \\[.5ex]
& HSGP4 & $0.73$ & $2.66$ & $12.49$ & $64.98$ & $303.71$ \\[1.5ex]
\multirow{2}{*}{$\omega$ (deg)} & SGP4 & $4.51$ & $4.51$ & $4.51$ & $-7.13$ & $-31.33$ \\[.5ex]
& HSGP4 & $2.59$ & $-5.26$ & $-6.39$ & $9.87$ & $40.43$ \\[1.5ex]
\multirow{2}{*}{$M$ (deg)} & SGP4 & $-4.57$ & $-4.57$ & $-4.57$ & $6.38$ & $27.52$ \\[.5ex]
& HSGP4 & $-2.59$ & $5.28$ & $6.41$ & $-9.76$ & $-40.22$ \\[.5ex]
\hline
\end{tabular}
\end{table}

\begin{figure}[!!htp]
\begin{subfigmatrix}{2}
\subfigure[Semimajor Axis.]{\includegraphics{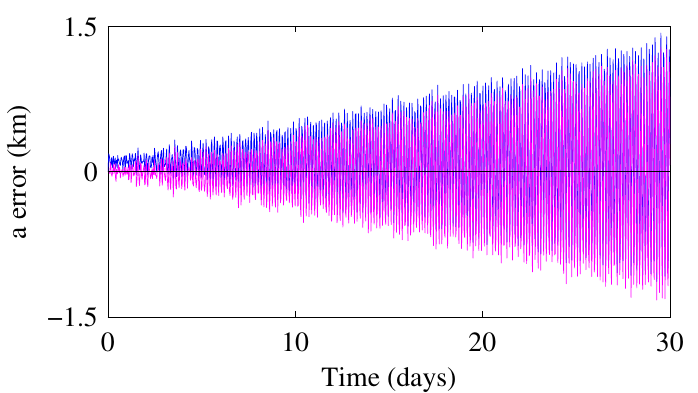}}
\subfigure[Right Ascension of Ascending Node.]{\includegraphics{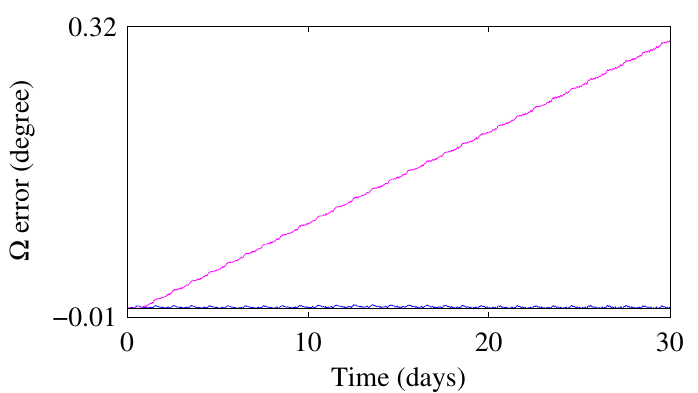}}
\subfigure[Eccentricity.]{\includegraphics{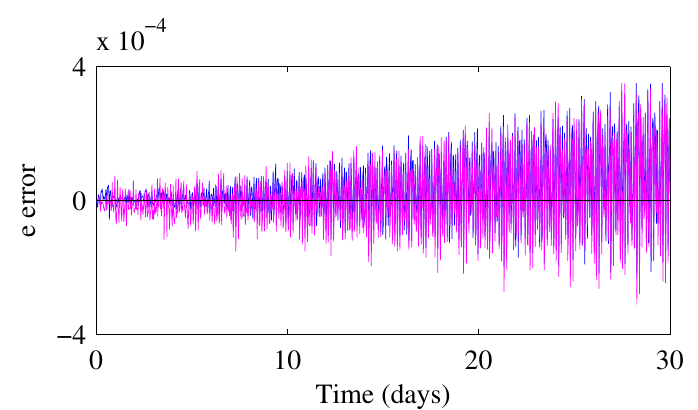}}
\subfigure[Argument of Perigee.]{\includegraphics{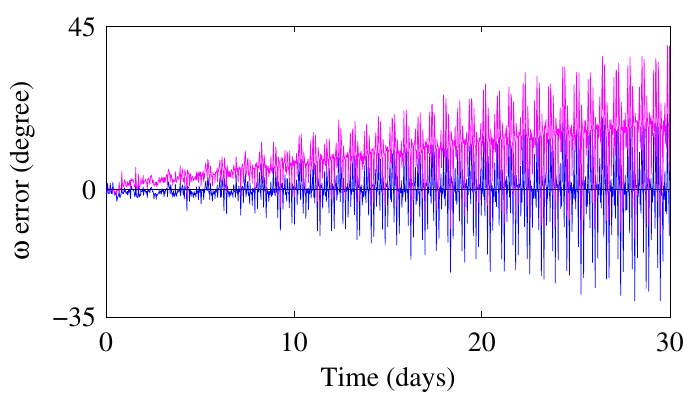}}
\subfigure[Inclination.]{\includegraphics{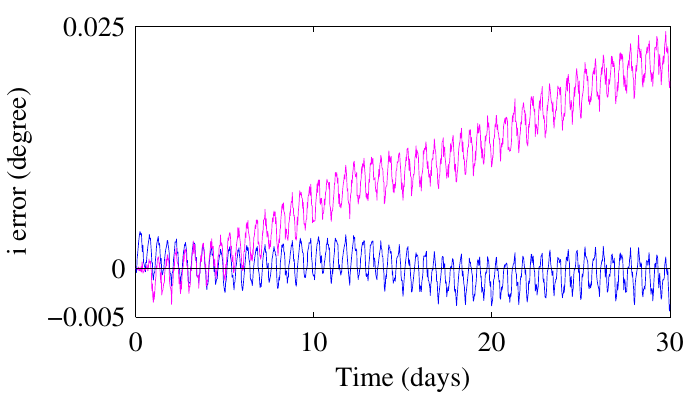}}
\subfigure[Mean Anomaly.]{\includegraphics{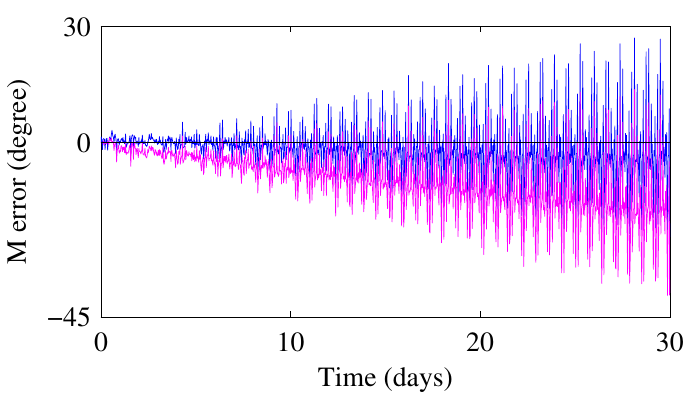}}
\label{}
\end{subfigmatrix}
\caption{Orbital element errors corresponding to SGP4 (blue) and HSGP4 (magenta), modeling $(l,g,h,L,G,H)$ Delaunay variables, during a 30-day propagation. Only 25$\bm{\%}$ of data have been plotted, without affecting the contour of the resulting figures, for the sake of clarity.}
\label{forb_lghLGH_30d}
\end{figure}

In order to gain some insight into the behavior of both propagators with respect to the orbital elements, Figure~\ref{forb_lghLGH_30d} depicts their errors during a $30$-day propagation.
The first remarkable fact confirms the conclusion previously drawn from Table~\ref{torb_lghLGH} concerning the qualitative behavior of the argument of the perigee and the mean anomaly, whose plots are practically symmetrical, and thereby show complementary errors. Although these errors are quite high, approximately $30^\circ$ in the case of SGP4 and $40^\circ$ in HSGP4 for a $30$-day propagation, the modeling of both angles, which correspond to the Delaunay variables $g$ and $l$, can lead to their mutual compensation.

On the other hand, the inclination and the right ascension of the ascending node plots confirm that HSGP4 behaves worse than SGP4, although their difference is very reduced, as can be deduced from Table~\ref{torb_lghLGH}: only $0.02^\circ$ in the case of the inclination, and $0.3^\circ$ for the right ascension of the ascending node after a $30$-day propagation. The modeling of the Delaunay variables directly associated with these orbital elements, $H$ in the case of the inclination, and $h$ as the right ascension of the ascending node, does not seem to be relevant for the accuracy of HSGP4 for the propagation time considered in this case.

Finally, the semimajor axis, related to $L$, and the eccentricity, connected with $G$, effectively show similar results for both propagators; therefore their contribution could be also irrelevant for the accuracy of the hybrid propagator in this case.

It is worth noting that the use of separate univariate forecasts converts the hybrid propagation methodology in a very flexible tool in the sense of adaptability, since it allows including only the effects that we are interested in modeling in the hybrid orbit propagator. In consequence, the application of this technique does not provide a single hybrid propagator, but an entire family of propagators in which the error models of different subsets of variables can be selected.

With the aim of illustrating this concept, and also in order to identify the most influential variables that lead to an accurate, though simple, hybrid orbit propagator, Table~\ref{tdist_max_subsets} compares the standard SGP4 propagator and three propagators of the family HSGP4, when the errors of three different subsets of Delaunay variables are modeled, in terms of their maximum position errors for several time spans. The three analyzed hybrid propagators are called HSGP4$_{(l,g)}$, which only considers the angles $(l,g)$, HSGP4$_{(l,g,L,G)}$, which takes into account the variables $(l,g,L,G)$, and, finally, HSGP4$_{(l,g,h,L,G,H)}$, which models the six Delaunay variables. It can be observed that the mean anomaly $l$ and the argument of the perigee $g$ are the variables whose error modeling contributes the most to the accuracy of the hybrid propagator. Modeling also their conjugate momenta, $L$ and $G$, does not have any appreciable effect, therefore they can be ignored for the sake of a simpler model. Similarly, the influence of the right ascension of the ascending node, $h$, and its conjugate momentum, $H$, is rather reduced; they can even worsen the distance error after a $30$-day propagation as a consequence of being long-period terms, whose slow dynamics can scarcely be modeled in a reduced control interval of only $0.7$ days. Therefore, modeling only $l$ and $g$ variables can be enough so as to implement the hybrid propagator.

\begin{table}[htbp!!]
%\fontsize{10}{10}\selectfont
\caption{Maximum position errors (km) of SGP4 versus three propagators of the family HSGP4, adjusted to different subsets of Delaunay variables, after several propagation spans.}
\label{tdist_max_subsets}
\centering 
\begin{tabular}{ccccc}
\hline
Propagation & \multirow{2}{*}{SGP4} & \multirow{2}{*}{HSGP4$_{(l,g)}$} & \multirow{2}{*}{HSGP4$_{(l,g,L,G)}$} & \multirow{2}{*}{HSGP4$_{(l,g,h,L,G,H)}$} \\
span & & & & \\
\hline
$0.7$ days & $\phantom{4}10.551$ & $\phantom{4}1.088$ & $\phantom{4}0.994$ & $\phantom{4}1.019$ \\[.5ex]
$\phantom{l}1$ day & $\phantom{4}14.235$ & $\phantom{4}3.206$ & $\phantom{4}3.195$ & $\phantom{4}3.198$ \\[.5ex]
$\phantom{3}2$ days & $\phantom{4}28.650$ & $\phantom{4}5.883$ & $\phantom{4}5.893$ & $\phantom{4}5.708$ \\[.5ex]
$\phantom{3}7$ days & $101.164$ & $20.477$ & $20.500$ & $19.483$ \\[.5ex]
$30$ days & $486.738$ & $41.100$ & $41.099$ & $42.969$ \\[.5ex]
\hline
\end{tabular}
\end{table}

\begin{figure}[!!htp]
\begin{subfigmatrix}{2}
\subfigure[Distance Error.]{\includegraphics{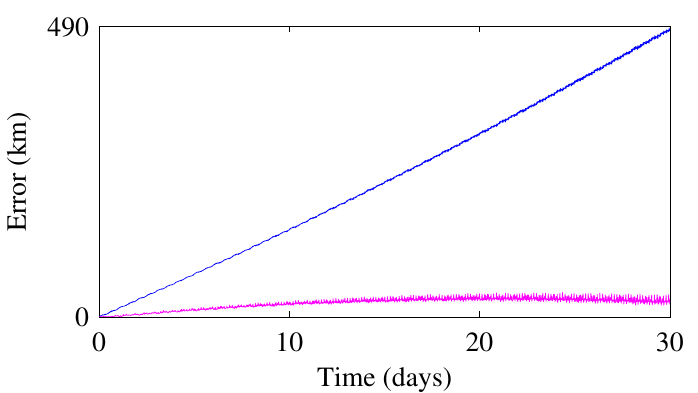}}
\subfigure[Along-Track Error.]{\includegraphics{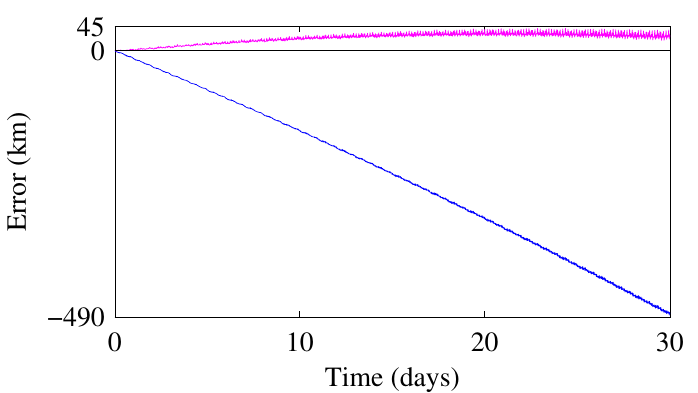}}
\subfigure[Cross-Track Error.]{\includegraphics{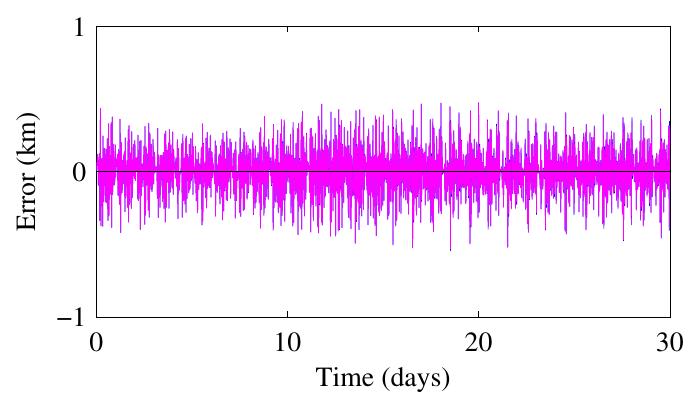}}
\subfigure[Radial Error.]{\includegraphics{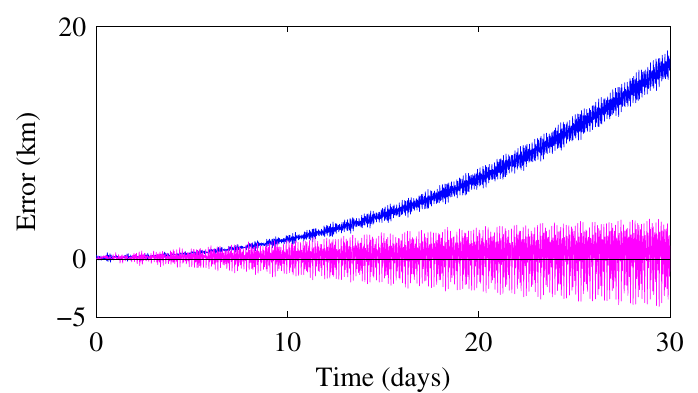}}
\label{}
\end{subfigmatrix}
\caption{Distance error, along-track error, cross-track error, and radial error corresponding to SGP4 (blue) and HSGP4$_{(l,g)}$ (magenta), modeling $(l,g)$ Delaunay variables, during a 30-day propagation. Both plots are virtually identical in the case of the cross-track error. Only 25$\bm{\%}$ of data have been plotted, without affecting the contour of the resulting figures, for the sake of clarity.}
\label{fdist_lg}
\end{figure}

Nevertheless, the most remarkable conclusion that can be drawn from Table~\ref{tdist_max_subsets} is that the hybrid propagation methodology clearly contributes to extend the validity of the corresponding TLE. In fact, the maximum position error is five times lower in the three considered HSGP4 propagators than in SGP4 after one week, and eleven times lower after $30$ days of propagation. The maximum position error of the HSGP4 propagators after $30$ days, which ranges from $41$ to $43$ km, is equivalent to the SGP4 error after only $3$ days. Figure~\ref{fdist_lg} depicts the distance error, as well as its three components in the tangent, normal, and binormal directions, for both the standard SGP4 propagator and HSGP4$_{(l,g)}$, which only models the $l$ and $g$ angles.

Both Table~\ref{tdist_max_subsets} and Figure~\ref{fdist_lg} show that the position error of HSGP4$_{(l,g)}$ remains one order of magnitude below the position error of the standard SGP4. Nevertheless, Figure~\ref{forb_lghLGH_30d} did not present such an improvement in any of the six orbital elements. However, a noticeable complementary behavior of the mean anomaly and the argument of the perigee errors was detected in those figures. Since both variables are the most influential in HSGP4, as previously deduced, and hence the only ones that are being modeled in Figure~\ref{fdist_lg}, it seems quite evident that their corresponding errors are being mutually compensated.

\begin{figure}[!!htp]
\centering
\includegraphics{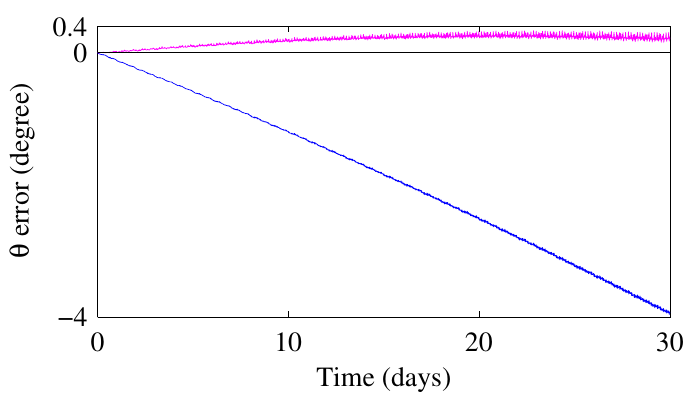}
\caption{Argument of latitude error corresponding to SGP4 (blue) and HSGP4$_{(l,g)}$ (magenta), modeling $(l,g)$ Delaunay variables, during a 30-day propagation. Only 25$\bm{\%}$ of data have been plotted, without affecting the contour of the resulting figure, for the sake of clarity.}
\label{fhill_lg_30d_th}
\end{figure}

In order to verify this assumption, we consider another variable that comprises both angles, the argument of the latitude $\theta$, which is defined as the addition of the true anomaly, $f$, whose behavior is similar to the mean anomaly, and the argument of the perigee, $\omega$. Therefore, the argument of the latitude collects the short-period terms, which are the most influential for short-term propagations, and also the easiest to be modeled during a reduced control interval of only $0.7$ days. Figure~\ref{fhill_lg_30d_th} shows the argument of the latitude error of both the standard SGP4 and HSGP4$_{(l,g)}$ propagators for a $30$-day propagation. As previously presumed, the error of HSGP4$_{(l,g)}$ remains one order of magnitude below the error of SGP4.

\begin{figure}[!!htp]
\centering
\includegraphics[width=\linewidth]{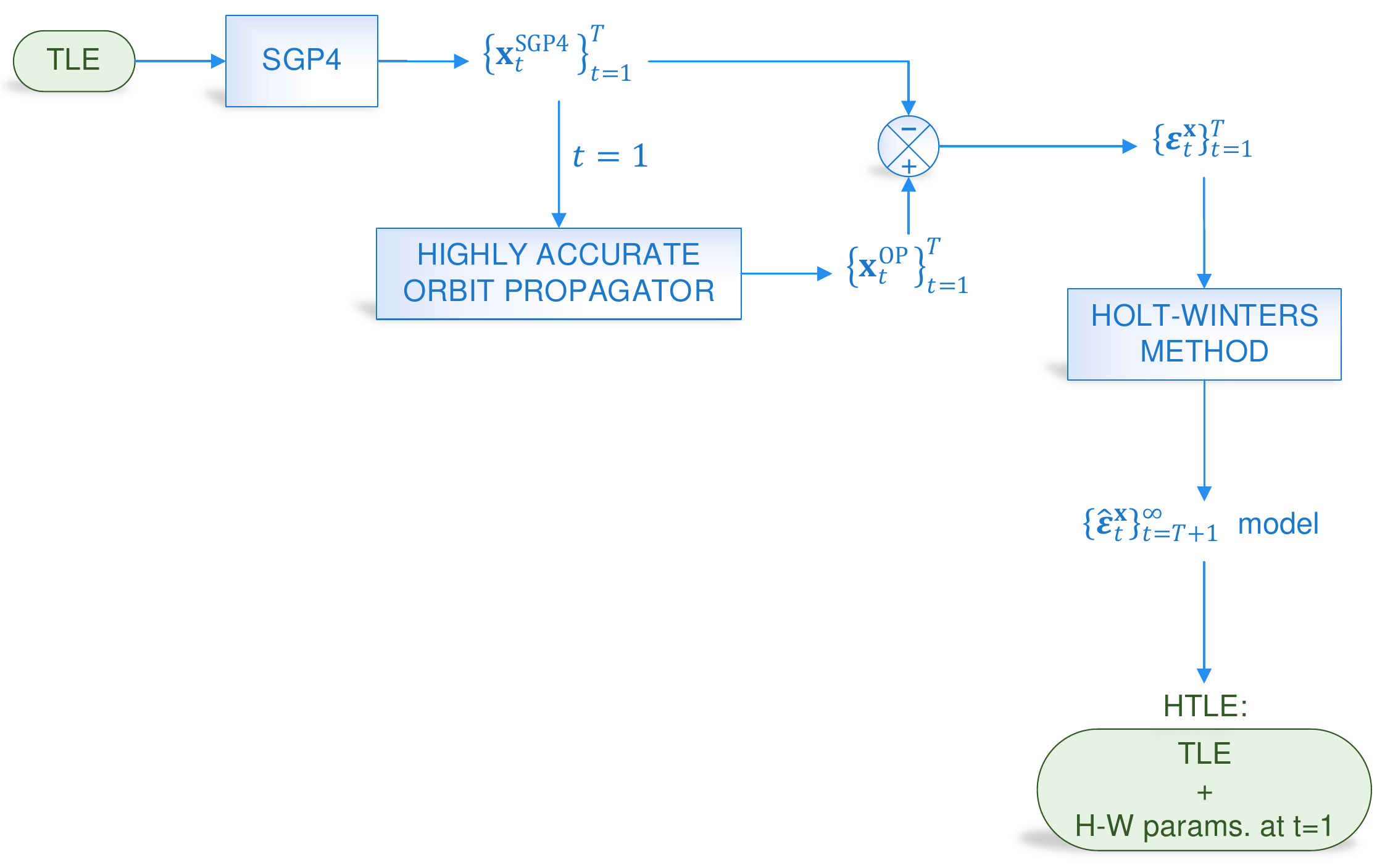}
\caption{Flow chart summarizing the process followed in the illustrative case presented in this study.}
\label{fflow_htle}
\end{figure}

Finally, Figure~\ref{fflow_htle} summarizes the main steps described in this section, in which we have presented an example in order to illustrate the application of the hybrid methodology to the particular case of the SGP4 propagator. Nevertheless, it is worth noting that, in this case of the hybrid HSGP4 propagator, the modeling process is not intended for the end user, but for the organization in charge of distributing the TLEs, which handles first-hand information that is not accessible for the end user, mainly real observations, which eliminates the need for the pseudo-observations that we have used for illustrative purposes. By following the proposed methodology, extended TLEs that include some additional information, that is, the Holt-Winters parameters of the error model, can be distributed, thus constituting what we have called hybrid TLEs or HTLEs.

For example, if the seasonal component is characterized with ten points, like in the case presented in this study, the number of additional parameters to be included in the corresponding HTLE is $12$: the ten points of the periodic oscillation plus the level and slope of the secular component. As the flow chart shows, the Holt-Winters parameters can even be referred to the initial instant $t_1$, instead of being associated to the end of the control interval $t_T$, for a greater simplicity of the process.

The end user can benefit from the extended validity of HTLEs through the use of the hybrid HSGP4 propagator, which comprises the standard SGP4 plus an error corrector. The calculation of the orbiter ephemerides requires the error corrector to deduce the argument of the latitude from the output of SGP4, to estimate the correction to be added from the HTLE additional parameters, according to line 11 in algorithm~\ref{algHW}, and finally to convert the corrected argument of the latitude, together with the rest of SGP4 output variables, into the desired set of coordinates.

It is worth noting that any other variable or subset of variables, apart from the argument of the latitude, can be chosen to have their errors modeled, in the HTLE-distributor side, and corrected, in the end-user side, hence allowing for maximum flexibility to choose the most influential variables for each orbital regime, which is an ongoing line of research.

\section{Conclusion}

The application of the hybrid propagation methodology to the well-known SGP4 orbit propagator allows for the extension of the validity of TLEs with little complexity and computational burden for the end user. A remarkable result can be achieved by modeling the argument of the latitude error, which implies handling just one time series, and thus a simplification of the model. In addition, as the argument of the latitude is one of the polar-nodal variables, it is non-singular, and therefore valid for small values of eccentricity and inclination.

Nevertheless, different variables or subsets of variables can be processed according to this methodology, which allows having a complete family of hybrid propagators, from which to choose the best suited for each orbital regime. This is an ongoing line of research in which valuable results are expected.

This methodology could be adopted in a very simple and non-intrusive way. It would require that TLEs be distributed with some additional information that the end user would need so as to complement the standard SGP4 output in order to correct its error. The institution in possession of the osculating elements would have to model the error of each orbiter, in order to incorporate the parameters deduced during the application of the Holt-Winters algorithm to the modeling of the error time series into the hybrid TLEs, HTLEs. The implementation of the correction is totally non-intrusive for SGP4. The end user would simply have to use a hybrid SGP4 propagator, HSGP4, composed of the standard SGP4 plus an error corrector.

It is worth mentioning that the overhead imposed by the proposed methodology is very reduced, not only concerning the amount of additional information to be included in HTLEs, which can be limited to just a few parameters, but also with regard to the computational burden, which implies a few basic operations, both in the HTLE-distributor and in the end-user sides.

\section*{Acknowledgments}

This work has been funded by the Spanish State Research Agency and the European Regional Development Fund under Projects ESP2014-57071-R and ESP2016-76585-R (AEI/ERDF, EU). The authors would like to thank Elecnor Deimos for having provided highly accurate ephemerides of the satellite Deimos~1. Valuable suggestions received from anonymous reviewers during the publishing process are also appreciated.

%\section*{References}

\bibliographystyle{elsarticle-num}
\bibliography{references}

\end{document}